\documentclass[aps,twocolumn,
rsi,%
 amsmath,amssymb,
preprint,%
 reprint,%
]{revtex4-1}

\usepackage{graphicx}
\usepackage[compact]{titlesec}         
\titlespacing{\section}{0pt}{0pt}{0pt} 
\AtBeginDocument{
  \setlength\abovedisplayskip{0pt}
  \setlength\belowdisplayskip{0pt}}
\usepackage[rawfloats=true]{floatrow} 
\restylefloat{figure}  
\usepackage{bm}
\usepackage{float}
\usepackage{color}

\begin{document}
\title{Stability analysis of a Boeing 737-800}
\author{Blessed Arthur Ngwenya $^{1,2}$ and Bob Osano$^{2,3}$ \\
\small{$^{1}$ Department of Physics,\\ University of Cape Town (UCT), Rondebosch 7701, Cape Town, South Africa\\}
\small{$^{2}$Cosmology and Gravity Group, Department of Mathematics and Applied Mathematics,\\ University of Cape Town (UCT), Rondebosch 7701, Cape Town, South Africa\\}
\small{$^{3}$Academic Development Programme,\\ University of Cape Town (UCT), Rondebosch 7701,Cape Town, South Africa}}

\email{NGWBLE001@myuct.ac.za} 
\email{bob.osano@uct.ac.za}
\date{\today} 

\maketitle

\section*{ABSTRACT}
The Anthena Vortex Lattice (AVL) \cite{avl} program describes aerodynamic and flight-dynamic analysis of rigid aircraft of arbitrary configuration and we use it to analyse the stability modes of a Boeing 737-800 under various small perturbations about trimmed equilibrium. We perturb the aircraft's free-stream velocity, its banking angle, its mass (which can be considered in conditions such as fuel dumping) and also analyse how it behaves at different heights (different air densities). We then compute the time it takes the various stability modes to return to equilibrium and show that this time obeys the logistic growth model for the Dutch roll and Short period mode when the velocity is perturbed and when varying the height above sea level of the aircraft.

\section{Introduction}
Analysing the behaviour of an aircraft requires us to set-up and solve the aircraft equations of motion\cite{Etkin_59,Etkin, Carpenter, Cunis_2019}. These take in different inputs from the various controls such as the aileron, elevator,  rudder and throttle, as well as different considerations like  the flight condition or atmospheric disturbances. Solving them  outputs various quantities of interest, like the displacement, velocity and acceleration of the aircraft. The dynamic  relationship between these input and output variables is described by aircraft response transfer functions \cite{transferfunctions}.

Solving the response transfer functions results in a characteristic polynomial which can be used to obtain the  stability modes of the aircraft. Given the difficulty in the computation of various quantities of interest, a number of approximations are made and the motion of the aircraft is restricted to small perturbations. However, various software has been developed to solve these equations and model aircraft behaviour. We use the Anthena Vortex Lattice method developed at MIT to analyse the stability modes of a Boeing 737-800 under small perturbations.

The Athena Vortex Lattice (AVL) is an aerodynamic analysis method which is based on the extended Vortex Lattice Method (VLM) \cite{paz2015introduction, karamcheti1966principles}. It was created by Mark Drela from MIT Aero and Astro as well as Harold Youngren and is described on \cite{avl}. Its use for aerodynamic analysis is discussed on \cite{budziak2015aerodynamic}. AVL can be used to develop aircraft configuration and perform aerodynamic analysis like dynamic stability analysis. The VLM is purely numerical and is based on solutions to Laplace's equation \cite{griffiths2005introduction} (i.e a vortex singularity as the solution of Laplace's equation). VLM calculates quantities such as induced drag, lift distribution for a given wing configuration etc.

\section{Equations of motion}
\begin{figure}[ht]
\begin{center}
\includegraphics[scale=0.40]{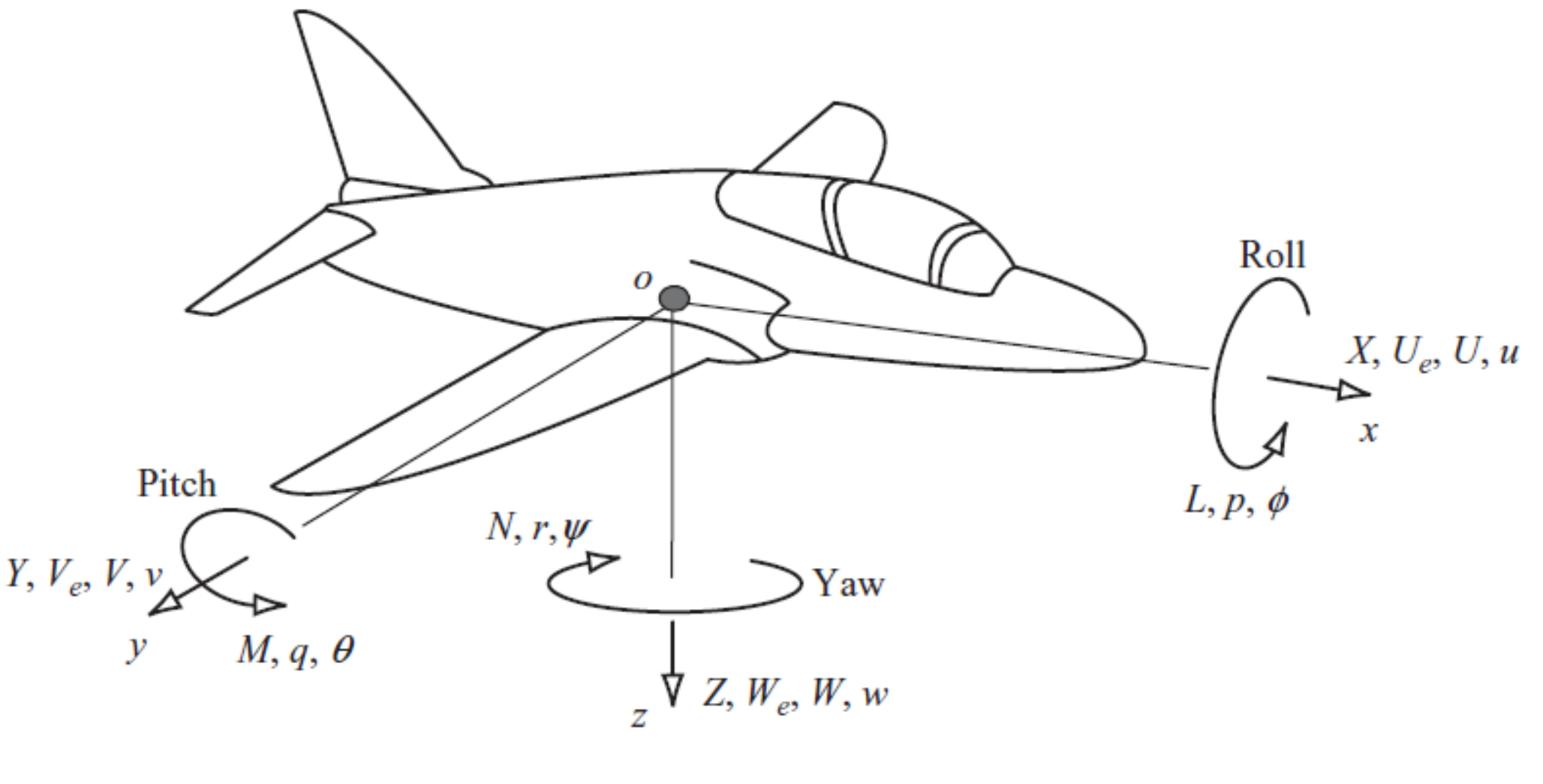}
\caption{\it Moving aircraft axes with motion variables \cite{main} of a generic fixed winged plane and not Boeing 737-800 whose stability we study in this article.}
\label{Fig1}
\end{center}
\end{figure}

Let $oxyz$ be a non-inertial coordinate system fixed to the body of the aircraft. If the origin $o$ represents the centre of mass of plane, axes are aligned so $x$-axis and $y$-axis are horizontal at equilibrium as shown in FIG (\ref{Fig1}) and will be referred to as stability axes. The departure of the aircraft from this orientation will allow for the study of disturbance from the steady reference flight condition. Stability of an aircraft is often studied under the stability theory, following the work of Lyapunov. It is known that for manned air crafts, instantaneous stability is not critical and as pointed out in \cite{Etkin} is neither a necessary nor a sufficient condition for successful flight. However, all air crafts will go through stable and unstable phases in their flight and the analysis of the time it takes to return to stability given a small perturbation from a steady state is crucial and may well
provide sufficient information about the general stability of the aircraft.

Initially, we assume trimmed equilibrium with steady velocity $V_0 = (U_e, V_e, W_e)$, meaning a steady/non-accelerating aircraft with the forces and moments acting on the air-frame balanced and summing up to zero. The perturbation variables are:

\begin{itemize}
    \item $X, Y, Z$ - the axial (drag), side and normal (lift) force
    \item $L, M, N$ - the rolling, pitching and yawing moment
    \item $p, q, r$ - the roll, pitch and yaw rate
    \item $U, V, W$ - the axial, lateral and normal velocity
    \item $\phi_{e}, \theta_{e}, \psi_{e}$ - equilibrium yaw, roll, pitch angles
\end{itemize}

The following figure shows the motion and perturbation variables. The linear quantities are positive when their direction is the same as the direction of action, while positive pitch is nose up, positive roll is right wing down and positive yaw is nose to the right as seen by the pilot.

The following table shows the motion variables for an aircraft in trimmed equilibrium and a perturbed one for each axis. All the components that are set to 0 are those that we'll ignore in our analysis of small perturbations.

\begin{table}[ht]
\centering
\begin{tabular}{c|c c c|c c c}\hline
	& \multicolumn{3}{c}{Trimmed equilibrium} &  \multicolumn{3}{|c}{Perturbed}\\
\hline
Aircraft axis & ox &oy &oz &ox &oy &oz\\
Force &0 &0 &0 &X &Y &Z\\
Moment &0 &0 &0 &L &M &N\\
Linear $\vec{v}$ &$U_e$ &$V_e$ &$W_e$ &U &V &W\\
Angular $\vec{v}$ &0 &0 &0 &p&q &r\\
Attitude &0 &$\theta_e$ &0 &$\phi$ &$\theta$ &$\psi$\\
\hline
\end{tabular}
\label{Table1}
\end{table}

In the case where we have transient perturbation with components (u,v,w), the velocity becomes: $U=U_e +u$, $V=V_e +v$ and $W=W_e +w$. Perturbation equations of motion can be set-up using Newton's 2nd law in each of the 6 degrees of freedom. It states $F=ma$ for the linear quantities and for the rotary quantities, (m) becomes the moment of inertia while (a) becomes the angular acceleration. We set up the equations at an arbitrary point P on the aircraft with coordinates(x,y,z). The local velocity and acceleration equations follow including linear and rotary terms.

\begin{align}
u &= \dot{x} -ry +qz	&a_x & = \dot{u} -rv +qw\\
v &= \dot{y} -pz +rx	&a_y & = \dot{v} -pw +ru\\
w &= \dot{z} -qx +py	&a_z & = \dot{w} -qu +pv
\end{align}

The equations we need for the 6 degrees of freedom are the generalised force equations (3 dof), including gravitational terms and the generalised moment equations (3 dof) given by Equations (4-9). Where X,Y and Z give the generalised forces in each direction, while L describes the rolling motion, M, the pitching motion and N the yawing motion. The gravitational terms represent the weight components in the steady state and $\theta _e$ is the angle between the horizon and $U_e$.

\begin{align}
m(\dot{U} -rV +qW)&= X 		&X_{ge}& = -mgsin\theta _e \\
m(\dot{V} -pW +rU)&= Y		&Y_{ge}& = 0\\
m(\dot{W} -qU +pV)&= Z		&Z_{ge}& = mgcos\theta _e
\end{align}

\begin{eqnarray}
L& =&I_x\dot{p}-(I_y -I_z)qr -I_{xz}(pq+\dot{r})\\
M& =&I_y\dot{q}-(I_x -I_z)pr +I_{xz}(p^2+r^2)\\
N& =&I_z\dot{r}-(I_x -I_y)pq +I_{xz}(qr+\dot{p})
\end{eqnarray}

In our analysis for small perturbations, we neglect terms involving products and squares of the velocity terms. We also assume that the aerodynamic force and moment terms in the moment equations are only dependant on the disturbed motion variables and their derivatives. We also neglect any longitudinal-lateral coupling, aerodynamic (or control) coupling derivatives.

Aerodynamic properties of an aircraft can be completely described using dimensionless parameters independent of air-frame geometry or flight condition. To make equations dimensionless, we divide by the appropriate force/moment parameter. We can also describe the equations of motion in state space form using state variables.

\section{Parameters and control/response module}
Aircraft response transfer functions are used to describe the dynamic relationships between the input and output variables through some mathematical models describing the dynamics of the aircraft. When we work with the decoupled equations of motion corresponding to small perturbations, longitudinal inputs correspond to longitudinal outputs and the same applies for the lateral. If $y(t)$ is the output which corresponds to the input $x(t)$, the transfer function $\mathcal{T}(s)$ is a ratio given by 

\begin{eqnarray}
\mathcal{T}_{yx}(s) &=&\frac{y(s)}{x(s)}
\end{eqnarray}

\begin{figure}[ht]
\begin{center}
\includegraphics[scale=0.35]{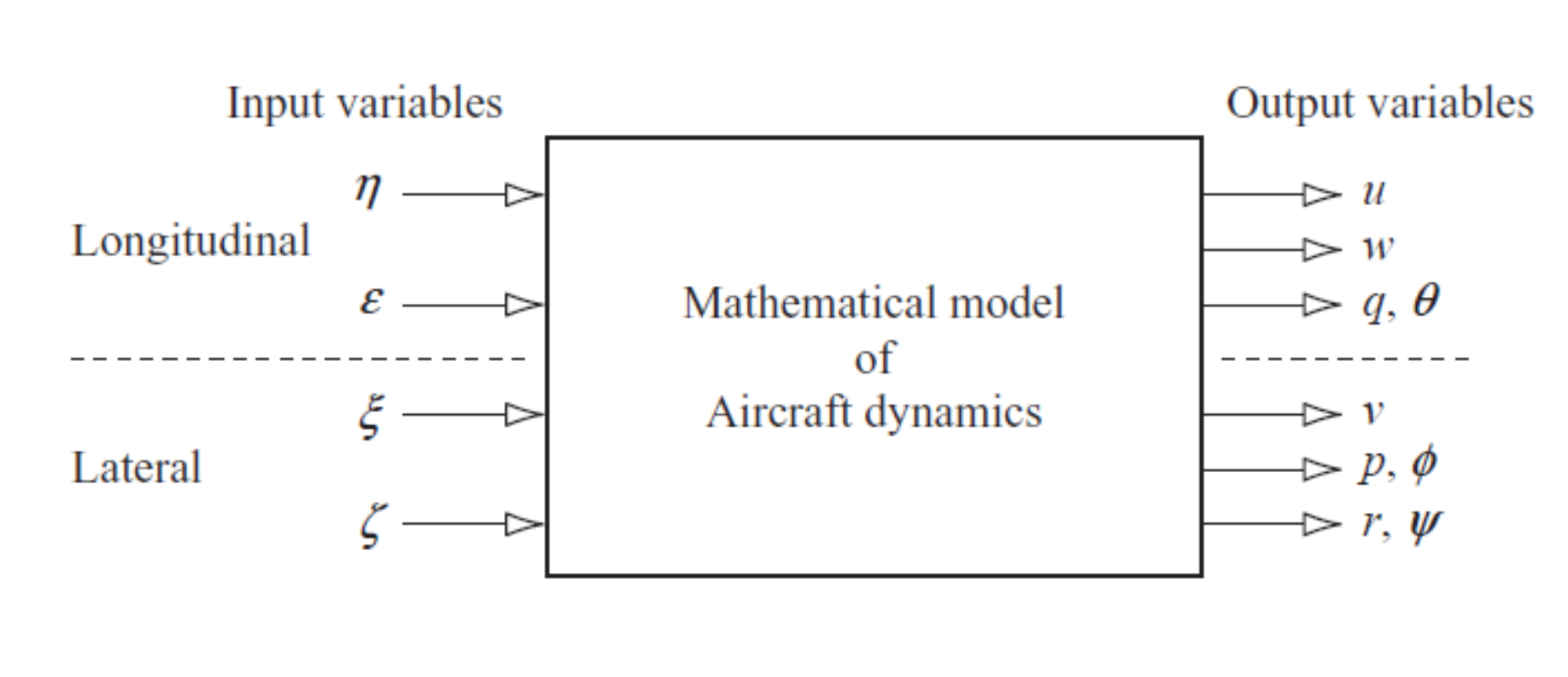}
\caption{\it Relationships of aircraft input and output variables \cite{main}. $\eta$ is the elevator angle perturbation, $\mathcal{E}$ is the throttle lever angle, $\xi$ is the aileron angle perturbation and $\zeta$ is the rudder angle perturbation. The output variables are the small perturbations as previously defined.}
\label{Fig2}
\end{center}
\end{figure}

We discuss next, specific cases of such transfer functions, and in particular the ones related to our study. Consider Fig (\ref{Fig2}) showing inputs and outputs classified as either longitudinal or lateral.

Transfer functions, as discussed above, are written as a ratio of two polynomials in the Laplace operator $s$. For example, a transfer function involving two lateral variables is found by perturbing the banking angle. The roll rate $p(s)$ response to the aileron $\xi(s)$ is given by:

\begin{eqnarray}
\mathcal{T}_{\vert_{p\xi}}(s)=\frac{p(s)}{\xi(s)}= \frac{{N_{\xi}^{p}}(s)}{\Delta(s)}
\end{eqnarray}
In this equation, $N_{\xi}^{p}(s)$ is a unique numerator polynomial in s relating roll response to elevator input and $\Delta(s)$ is the characteristic polynomial common for all lateral response transfer functions and gives the characteristic polynomial when equated to zero, which can be used to analyse the stability modes of the aircraft.

Laplace transforms play an important role in the analysis of an aircraft's state parameter and are worth reviewing. If $x(t)$ is any generic function of state, The transformation of its derivatives take the form;
\begin{eqnarray}
\mathcal{L}[\dot{x}]&=&\int_{0}^{\infty}e^{st}\dot{x}dt
=xe^{-st}|^{\infty}_{t=0}+s\int_{0}^{\infty}e^{-st}xdt
\end{eqnarray} where $\dot{x}=dx/dt$ and where we have used integration by parts given the conditions $xe^{-st}\rightarrow 0$ as $t\rightarrow 0.$ Similarly, the transformation of the $\ddot{x}$ can be found in a straight forward manner. We note that for the variables that we consider, the transformation processes are invertible. If $\dot{x}$ and $\ddot{x}$ are velocity and acceleration respectively, the corresponding transformations are given in short-hand notation by
\begin{eqnarray}
\mathcal{L}\{\dot{x}(t)\} &=& sx(s) -x(0)\\
\mathcal{L}\{\ddot{x}(t)\} &=& s^2x(s) -sx(0) -\dot{x}(0)
\end{eqnarray}

In these equations, $x(0)$ and $\dot{x}(0)$ give the initial values at time $t=0$. We then take the Laplace transform of the longitudinal/lateral equations of motion for small perturbations and express the result in matrix form. Then apply Cramer's rule to obtain the longitudinal/lateral response transfer functions, setting the characteristic polynomial to zero gives us the stability modes. This is discussed in detail on \cite{main, Etkin}. We now discuss stability analysis given slight departure from these modes. In this case, the stability of such a linear system, is determined by the roots of the characteristic polynomial. In particular, a mode will be convergent if the real part is negative which would indicate stability. A positive real part leads to divergence and hence instability. The technique so far discussed is for a linear system which is of the form $$\dot{\bf{x}}=f(\bf{x})$$ where $\bf{x}$ is a system of variables. One might wonder about how to handle a nonlinear system. For small disturbances such as the ones we consider, the method of linearization would suffice. In particular, if the nonlinear system of coupled variables is given by
\begin{eqnarray}
\dot{x}=f(x,y)\\
\dot{y}=g(x,y),
\end{eqnarray}
if ($x^{*}$,$y^{*}$) denote the values at equilibrium, i.e. 
\begin{equation}
  f(x^{*},y^{*})=0=g(x^{*},y^{*}) 
\end{equation}  then a small disturbance may be characterized\cite{Strogatz} as follows 
\begin{eqnarray}
\delta{x}=x-x^{*}\\
\delta{y}=y-y^{*}.
\end{eqnarray}
It is trivially true, from a Taylor series expansion, that 
\begin{eqnarray}
\dot{\delta x}&=&\delta x \frac{\partial f}{\partial x}+ \delta y \frac{\partial f}{\partial y}+\mathcal{O}(\delta x^{2},\delta x^{2}, \delta x\delta y)\\
\dot{\delta y}&=&\delta x \frac{\partial g}{\partial x}+ \delta y \frac{\partial g}{\partial y}+\mathcal{O}(\delta x^{2},\delta x^{2}, \delta x\delta y)
\end{eqnarray} $\mathcal{O}$ represents quadratic terms in $\delta x$ and $\delta y.$ It is clear that linearization, i.e. dropping of these higher order terms will lead to a linear system in $\delta x$ and $\delta y$ of the form\\

\begin{eqnarray}
\left(\begin{array}{c}
   \dot{\delta x} \\
   \dot{\delta y}  
\end{array}\right)=\left[\begin{array}{cc}
   \frac{\partial f}{\partial x}  &\frac{\partial f}{\partial y}  \\
   \frac{\partial g}{\partial x}  & \frac{\partial g}{\partial x}
\end{array}\right]_{_{(x^{*}, y^{*})}}\left(\begin{array}{c}
 {\delta x} \\
  {\delta y}  
\end{array}\right)
\end{eqnarray}

when evaluated at the equilibrium point $(x^{*}, y^{*})$.This suggests that one can obtain the equivalent Laplace transform of the Taylor series expansion about $(x^{*}, y^{*})$ up to linear order in the perturbation variables $\delta x, \delta y$. The resulting matrix is determined by the partial derivatives at $(x^{*}, y^{*})$. We define the following constants:
\begin{eqnarray}
\alpha_{1}&=&\frac{\partial f (x,y)}{\partial x}|_{(x^{*},y^{*})}\\
\alpha_{2}&=&\frac{\partial f (x,y)}{\partial y}|_{(x^{*},y^{*})}\\
\beta_{1}&=&\frac{\partial g (x,y)}{\partial x}|_{(x^{*},y^{*})}\\
\beta_{2}&=&\frac{\partial g (x,y)}{\partial x}|_{(x^{*},y^{*})}.
\end{eqnarray} The system of disturbance variables linearized about  $(x^{*}, y^{*})$ can be written as follows:
\begin{eqnarray}
\dot{\delta x}=\alpha_{1}\delta x+\alpha_{2}\delta y\\
\dot{\delta y}=\beta_{1}\delta x+\beta_{2}\delta y,
\end{eqnarray}
each of which may now be subjected to Laplace transformation as desired. It is easy to show that this yields
\begin{eqnarray}
{\bf A}\left(\begin{array}{c}
 {\delta x(s)} \\
  {\delta y(s)}  
\end{array}\right)=\left(\begin{array}{c}
   {\delta x(0)} \\
   {\delta y(0)}  
\end{array}\right)
\end{eqnarray}
where 
\begin{eqnarray}
{\bf A}=\left[\begin{array}{cl}
  s-\alpha_{1} &~~~~-\alpha_{2}\\
-\beta_{1}  &~~~~ s -\beta_{2}
\end{array}\right]
\end{eqnarray}

This system may now be analyzed as in the linear case by finding and analysing the characteristic polynomial. We note that 
\begin{eqnarray}
|{\bf A}|=f(s)=s^{2}-s(\alpha_{1}+\beta_{2})+(\alpha_{1}\beta_{2}-\alpha_{2}\beta_{1}).
\end{eqnarray} However, caution is advised as the success of the linearization is depended on the nature of the equilibrium point $(x^{*}, y^{*})$ \cite{Andronov}. We will not belabour this point.
With these two approaches we can now look at the stability modes for Boeing 737-800. The flight dynamics of 747-200 is analysed in \cite{Ogunwa_2016, Park_2017}.

\section{Stability Modes of a Boeing 737-800}
The characteristic equation for lateral motion commonly factorises into a pair of real and complex roots. The first real root describes the non-oscillatory Spiral mode which develops slowly and involves complex coupling in roll, yaw and sideslip. It's excited by a disturbance in sideslip, resulting in lift and eventually a yawing moment which turns the aircraft. The yawing motion causes a disturbance in roll so the aircraft flies a slowly diverging path in both roll and yaw resulting in an unstable spiral descent flight.

The second real root describes the Roll subsidence mode which has an exponential lag characteristic in the rolling motion. It rolls with an angular acceleration as described by Newton's 2nd law and the roll rate builds up exponentially until the restoring moment balances the disturbing motion and a steady roll is established.

The two complex roots describe the Dutch roll mode, which is a classical damped motion in yaw that couples into roll and sideslip. It is comparable to the Short period mode due to their similar magnitude in frequency and is a complex interaction between all three lateral degrees of freedom. This mode results in the aircraft rolling from side to side in some oscillatory cycle which eventually drops to zero.

\begin{figure}[ht]
\begin{center}
\includegraphics[scale=0.35]{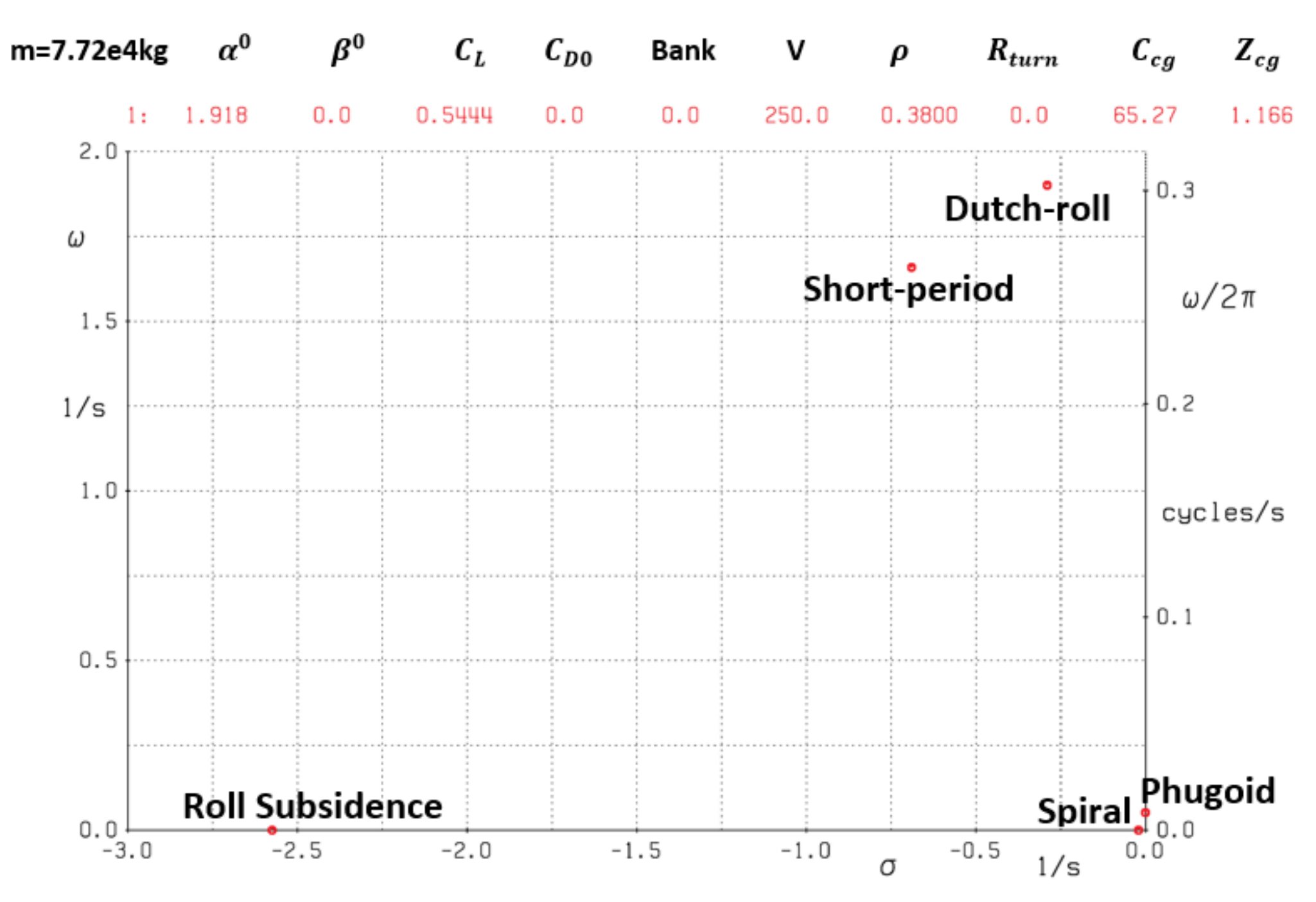}
\caption{\it Boeing 737-800 stability modes on the s-plane and excluding duplicates of roots lying on the negative half of the complex plane as obtained from AVL}
\label{fig3}
\end{center}
\end{figure}

The longitudinal motion characteristic polynomial is of 4th order and factorises into two repeated roots. The first pair of roots describes Phugoid mode, which is a classical damped harmonic motion and results in the aircraft flying in a gentle sinusoidal path about the trimmed height. The motion can be approximated by undamped harmonic motion conserving mechanical energy. The last pair of roots describe the Short-period mode which is a damped classical 2nd order oscillation in pitch. FIG (\ref{fig3}) shows the position of the roots on the s-plane for a Boeing 737-800.


\section{The Logistic Growth Model}
Analysis of nonlinear dynamical systems such as those involving an aircraft are complex and not easy to carry out without significant simplifying assumptions. It turns out that some biological systems display characteristics that are analogous to these mechanical systems but with the advantage that these biological systems are well studied and understood. For example, the flight stability of a hovering bumblebee was studied in \cite{Sun} which could improve our understanding of the stability of non-fixed winged vehicles. In our case, we draw analogy with a population growth model.

Various models are used to model growth of biological systems and these variously address population dynamics. The Verhulst logistic equation is used to approximate population growth and is given by Equation (32), its solution is given by Equation (33).

\begin{eqnarray}
\frac{dN}{dt} &=& rN\left(1-\frac{N}{K} \right)\\
N(t) &=& \frac{KN_0}{(K-N_0)e^{-rt} +N_0}
\end{eqnarray}

The model has the following key features:
\begin{itemize}
    \item r is the intrinsic decay rate and represents decay rate per capita
    \item $N_0$ is the population size at $t=0$ and the carrying capacity is $\lim_{x\to\infty} = K$
    \item The relative growth rate, $\frac{1}{N} \frac{dN}{dt}$ declines linearly with increasing population size
    \item The population at the point where growth rate is maximum (inflection point) is given by $N_{inf} = \frac{K}{2}$
\end{itemize}

\section{Results}
We can use the solution of the Verhulst logistic equation to model the time taken to reach equilibrium for various stability modes of a Boeing 737-800. In this case, we take $N(t)$ to be the time taken to re-establish equilibrium and $t$ to be the various quantities we perturb (i.e velocity).

\begin{figure}[ht]
\begin{center}
\includegraphics[scale=0.3,angle=-90,origin=c]{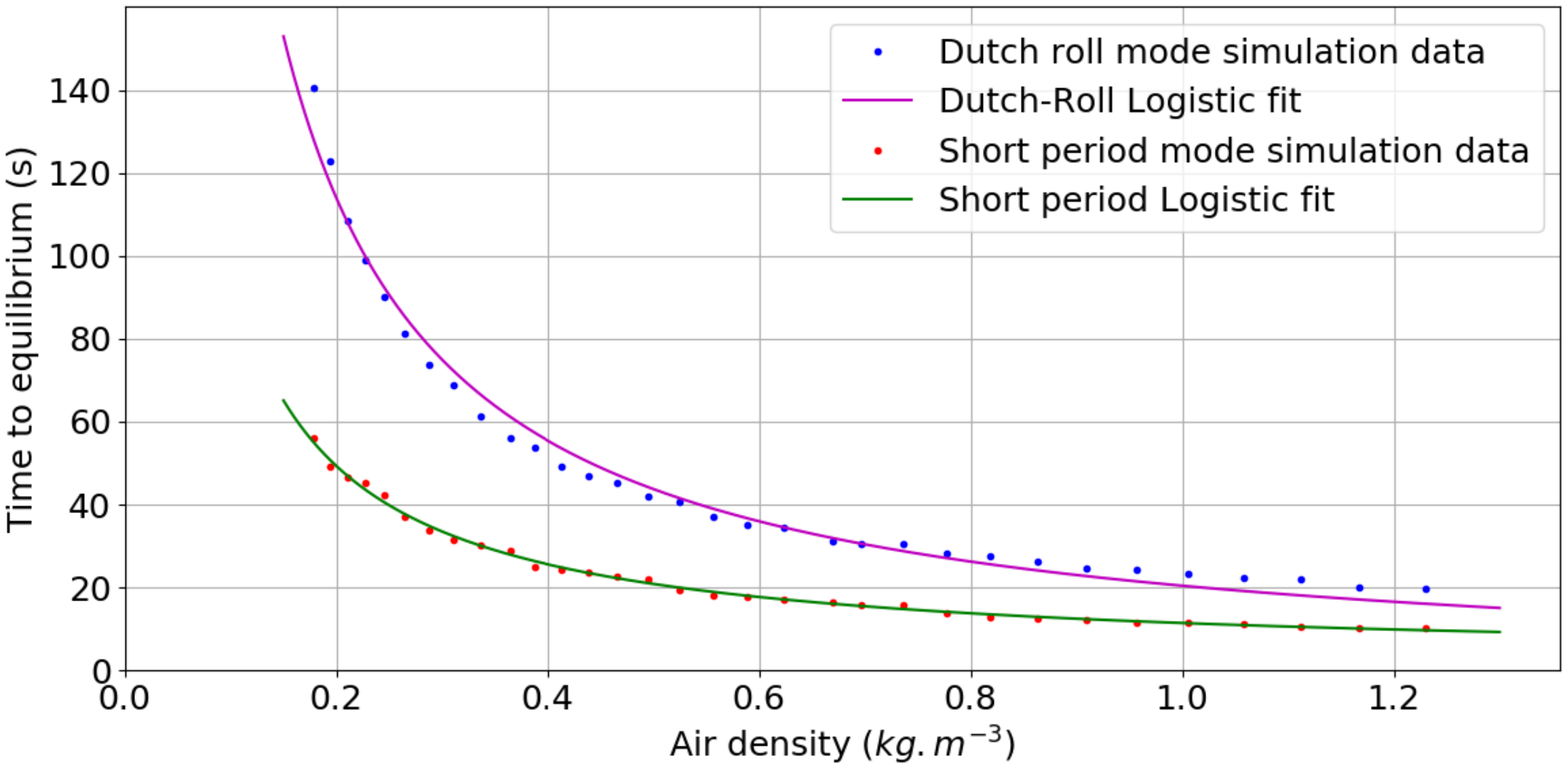}
\caption{\it Time taken to establish equilibrium as a function of air density (corresponds to different heights) for the Short period and Dutch roll modes of a Boeing 737-800}
\label{fig4}
\end{center}
\end{figure}

FIG (\ref{fig4}) shows the time taken to establish equilibrium as a function of air density (corresponding to different heights) for the Dutch roll and Short period mode, including the corresponding logistic fits. The Dutch-Roll mode has fit parameters: $N_0=1.03 \times 10^9$, $r=-2.69 \times 10^{-1}$ and $K = -6.29$. On the other hand, the Short-Period mode has fit parameters: $N_0=1.37 \times 10^8$, $r=3.79 \times 10^{-1}$ and $K = 3.60$.\\

\begin{figure}[ht]
\begin{center}
\includegraphics[scale=0.3,angle=-90,origin=c]{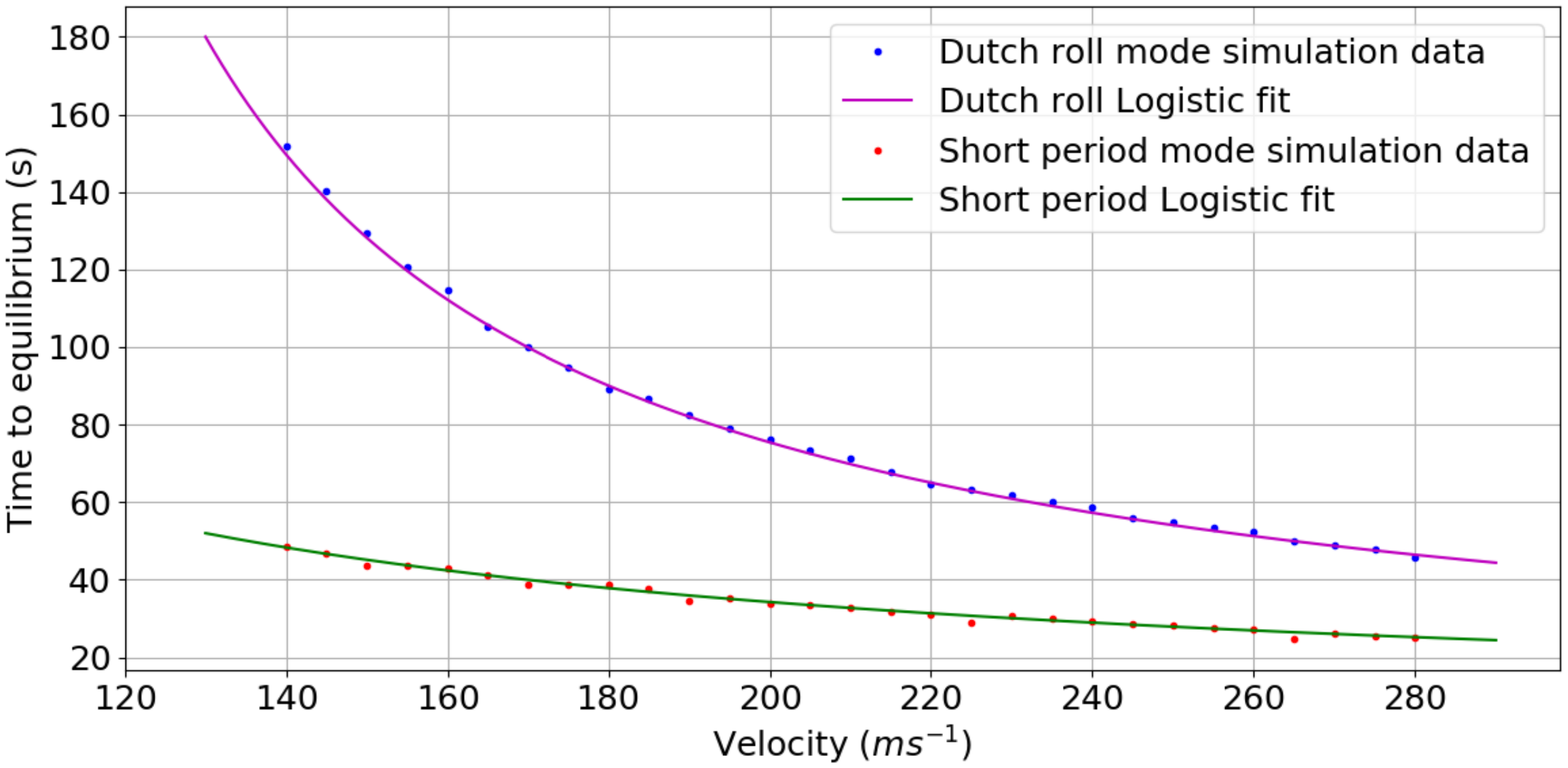}
\caption{\it Time taken to equilibrium as a function of velocity for the Short period and Dutch roll modes of a Boeing 737-800}
\label{fig5}
\end{center}
\end{figure}

We perform a similar analysis to find the time to equilibrium dependence on the velocity, the results are shown in FIG (\ref{fig5}). The AVL linearization assumes small perturbations, thus not completely valid for velocity perturbations that are large from the free-stream velocity.  The Dutch-Roll mode has fit parameters: $N_0 = 180.0$, $r=6.0 \times 10^{-3}$ and $K=7.10$. While the Short-Period mode has fit parameters: $N_0 = 52.0$, $r=8.0 \times 10^{-3}$ and $K=6.5$.\\

As can be seen from FIG (\ref{fig4}) and (\ref{fig5}), the time taken to establish equilibrium shows a damping system. This can be explained by considering the fact that both of the stability modes described manifest as a damped classical oscillator.

\begin{figure}[ht]
\begin{center}
\includegraphics[scale=0.3,angle=-90,origin=c]{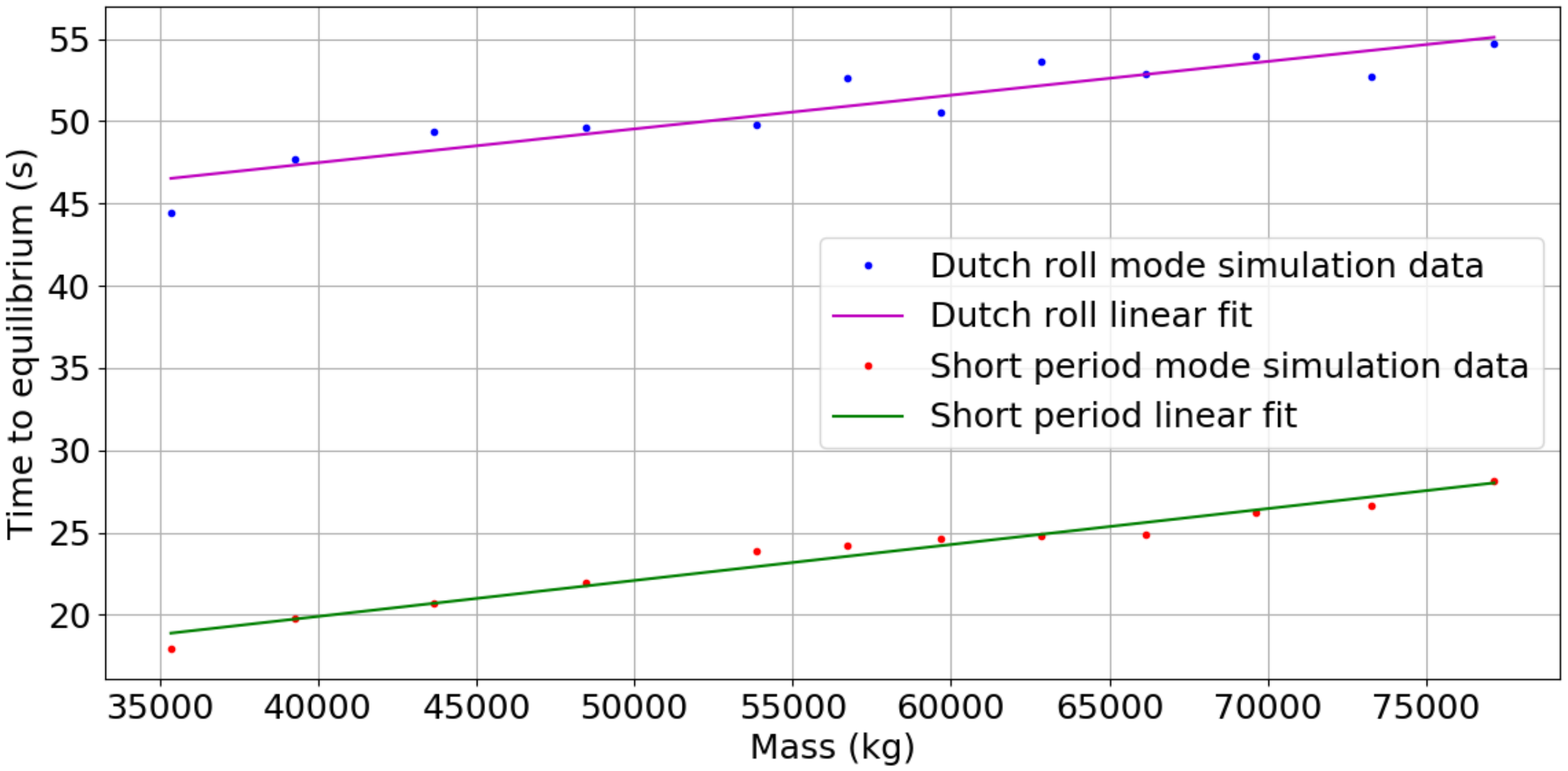}
\caption{\it Time taken to equilibrium as a function of the aircraft's mass for the Short period and Dutch roll modes of a Boeing 737-800}
\label{fig6}
\end{center}
\end{figure}

In some emergency situations, an aircraft may be forced to dump its fuel before landing in order to avoid structural damage, given that the structural landing weight is less than the structural take-off weight. FIG (\ref{fig6}) shows that a lighter aircraft will return to equilibrium faster than a heavier one for the Dutch roll and Short period modes. We also see that the time taken to establish equilibrium varies linearly with mass (m) and can be described by the linear fit Equation (34) for the Dutch roll mode and (55) for the Short period mode.

\begin{eqnarray}
T_{DR}(m) &=& 2.051 \times 10^{-4}m + 39.28\\
T_{SP}(m) &=& 2.184 \times 10^{-4}m + 11.16
\end{eqnarray}

\begin{figure}[H]
\begin{center}
\includegraphics[scale=0.3,angle=-90,origin=c]{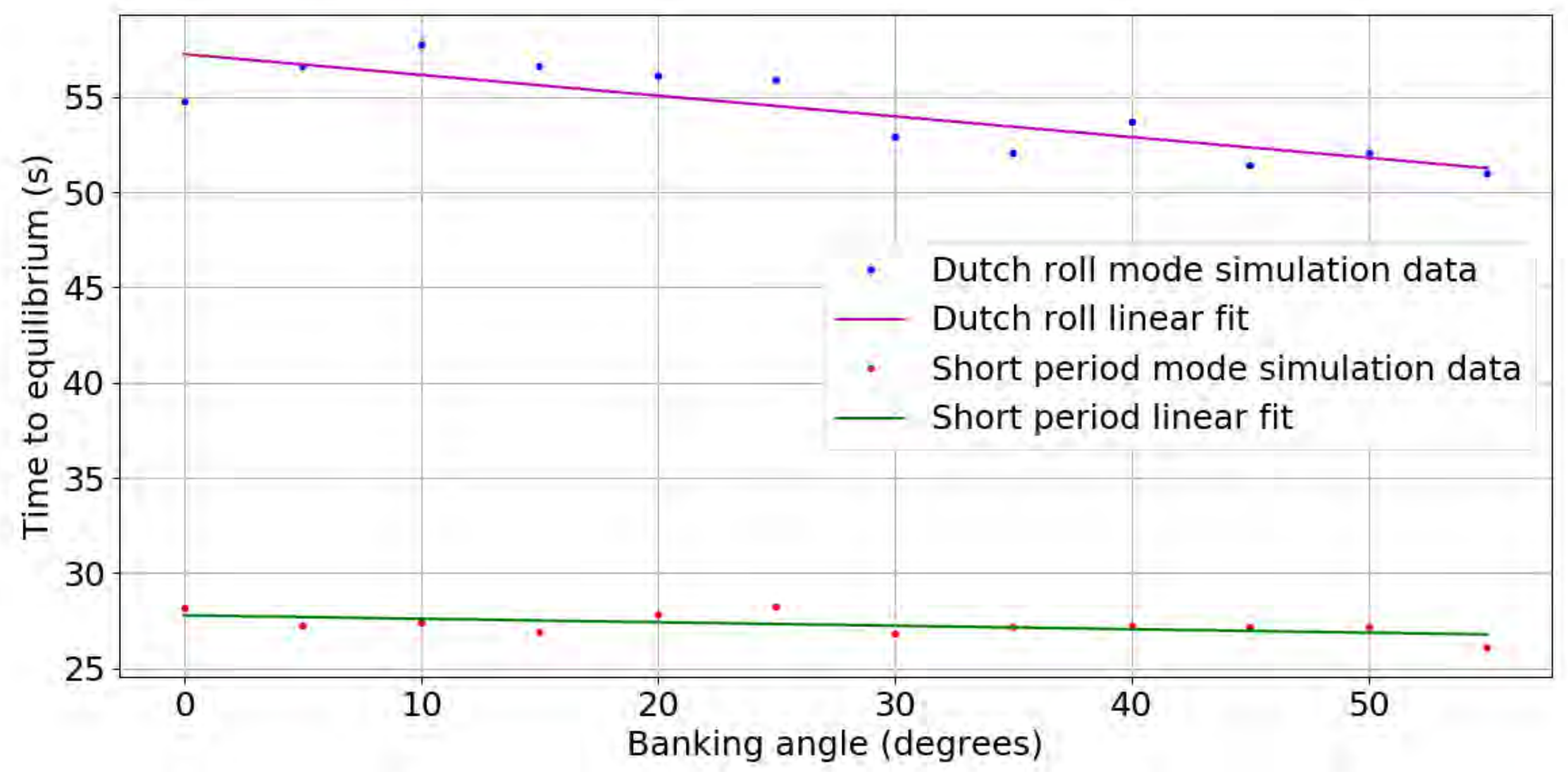}
\caption{\it{Time taken to equilibrium as a function of the banking angle for the Short period and Dutch roll modes of a Boeing 737-800}}
\label{fig7}
\end{center}
\end{figure}

When an aircraft turns, it banks the wings (rolls around the roll axis) at a particular angle in the direction of the desired turn. Once the bank angle is non-zero, then the wings are not level and the aircraft is not trimmed.  The turns are classified as shallow for a bank angle less than $20^{\circ}$, medium for $20^{\circ} \leq \text{bank angle} \leq 45^{\circ}$ and steep for bank angles greater than $45^{\circ}$. We also look at how the time taken to establish equilibrium changes with respect to the banking angle ($\phi$). FIG (\ref{fig7}) shows that the time to equilibrium is slightly higher for small bank angles and decreases linearly by a small amount as the banking angle is increased. Equations (36) and (37) describe the linear fit of the time to equilibrium for the Dutch roll and Short period mode respectively.

\begin{eqnarray}
T_{DR}(\phi) &=& -1.089 \times 10^{-1}\phi + 57.25\\
T_{SP}(\phi) &=& -1.829 \times 10^{-2}\phi + 27.78
\end{eqnarray}

One way to understand the dynamics studied here is to consider logarithmic decrement parameter 
\begin{eqnarray}
\Delta =\frac{1}{n}\log\left[\frac{\delta x(t_{n})}{\delta x(t_{n+1})}\right]
\end{eqnarray}for a time-discretized sampling of the perturbation variable $\delta x(t_{n})$ where time to equilibrium is given by $t_{eq}=(\sum n) T$  and where $\sum n$ is the total number of periods, $T$, it take to restore equilibrium.

\begin{figure}[ht]
\label{FigSampling}
\begin{center}
\includegraphics[scale=1.0]{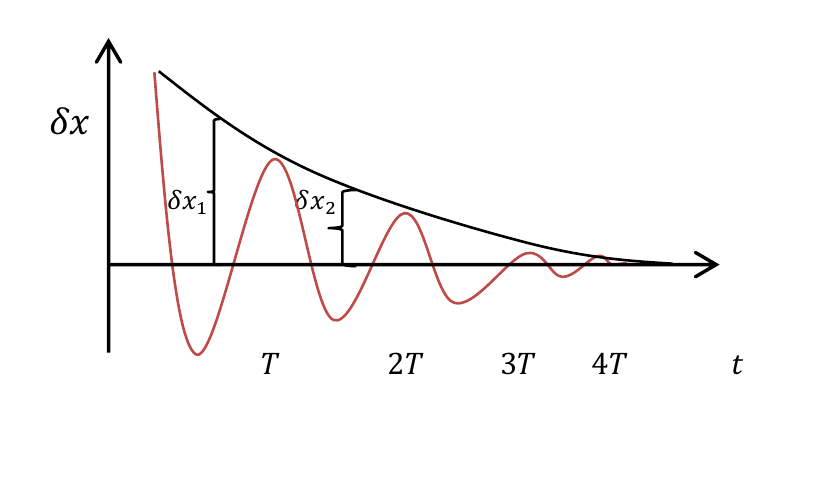}
\caption{\it Period-sampled logarithmic decay function}
\label{fig8}
\end{center}
\end{figure}

\begin{figure}[ht]
\begin{center}
\includegraphics[scale=0.35]{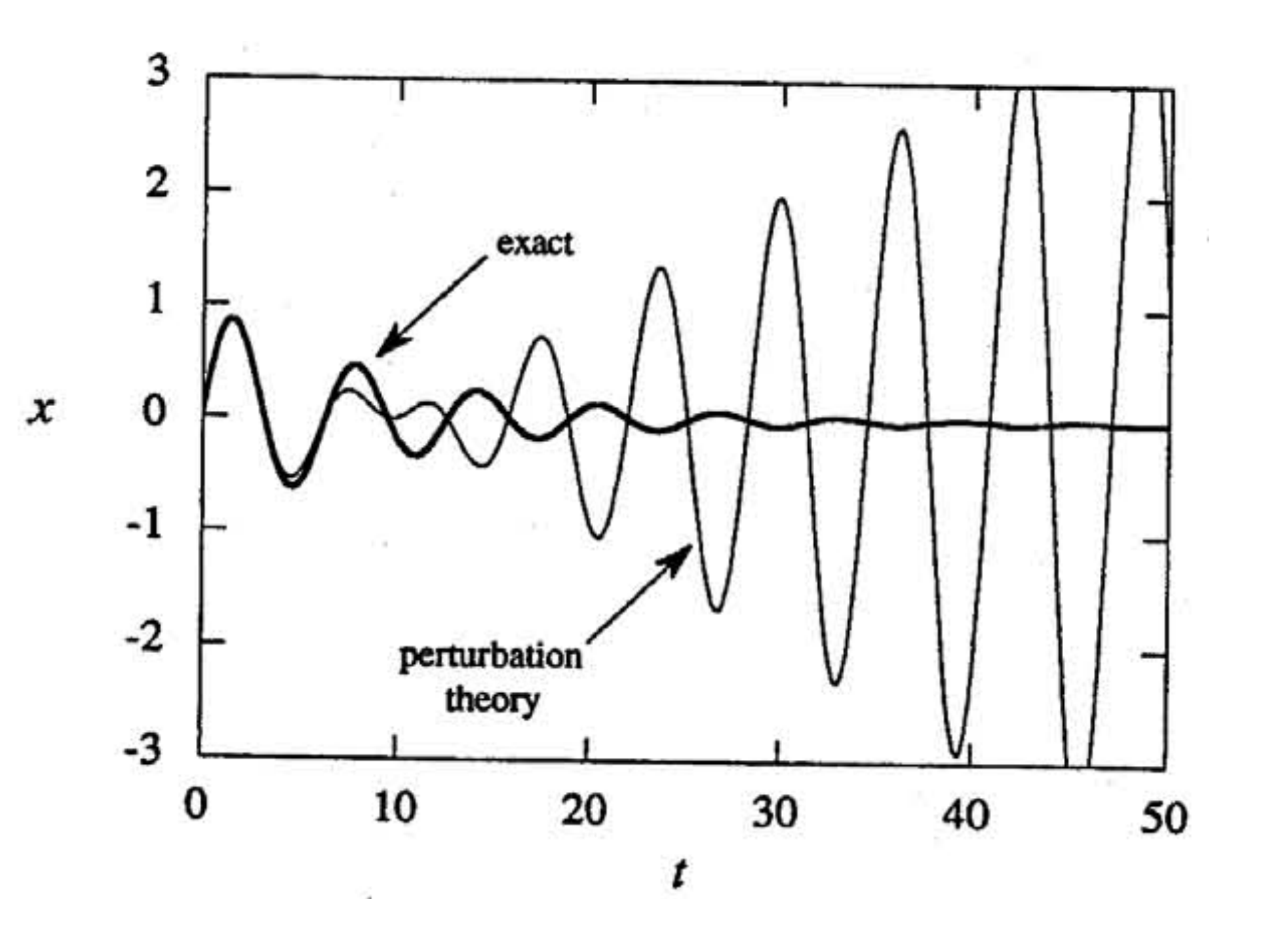}
\caption{\it The existence of two time scales when exact solution is compared to a perturbed approximation\cite{Strogatz}}
\label{fig9}
\end{center}
\end{figure}

In FIG (\ref{fig8}) we have fitted a periodic function such that subsequent peaks or crests of the oscillating function are joined by exponential segments which together coincide that of our decaying parameter. A simple measure of stability is the comparison of $\Delta$s for different graphs of the same period taken at the same time-step $n$. It follows that a comparative larger value of $\Delta$ is indicative of greater stability. All cases studied above are amenable to the $\Delta$ comparison technique. FIG (\ref{fig9}) is unique in that it shows growth rather than decay but it too can be $\Delta$-analysed and interpreted accordingly. We have used a logistic function to examine stability. It will be noted that one could use weakly nonlinear oscillators \cite{Strogatz}, but one must be careful because not all linearized system correctly approximate the exact solution. This is due to existence of two-time scales for weakly nonlinear oscillators. This is illustrated below in FIG (\ref{fig9}). The remedy is to apply the two-timing analytical technique and the use this for sampling the function of interest\cite{BB}.  

\section{Discussion and Conclusion}
The stability analysis of flight dynamics is not a new topic, the analysis of the stability of the multi-factorial and coupled dynamical systems has remained a challenge. In this study we set out to examine the stability of a Boeing 737-800 by examining how the vehicle responds to small disturbances. We have used both analytical and simulation techniques. We have devised a new periodic-sampling technique where an oscillating function is fitted to match the function of interest and this is used to indirectly quantify the stability given the small disturbances. The counter-check to this method was the simulation using the publicly available AVL code.

In our analysis of the Dutch roll and Short period modes of a Boeing 737-800, we have shown that the time taken to return to equilibrium is longer at high altitudes (low air densities) and shorter at low altitudes (high air densities). This time taken to return to equilibrium decays according to a logistic growth model with decreasing altitude. This conclusion is confirmed by the $\Delta-$analysis method we have developed. We have also shown that a lighter Boeing 737-800 re-establishes equilibrium faster than a heavier one and that it re-establishes equilibrium slightly quicker if it is banked than when trimmed.

\section{Acknowledgements}

Both authors thank the University of Cape Town (UCT) and the  South African National Research Foundation (NRF) for funding. B. A Ngwenya acknowledges Valumax Projects (Pty) Ltd and the University of Cape Town for their generous financial support towards this work.  B. Osano is thankful for funding from the University of Cape Town Next Generation Professoriate (NGP) programme.

\section{References}
\bibliography{references}

\begin{thebibliography}{17}%
\makeatletter
\providecommand \@ifxundefined [1]{%
 \@ifx{#1\undefined}
}%
\providecommand \@ifnum [1]{%
 \ifnum #1\expandafter \@firstoftwo
 \else \expandafter \@secondoftwo
 \fi
}%
\providecommand \@ifx [1]{%
 \ifx #1\expandafter \@firstoftwo
 \else \expandafter \@secondoftwo
 \fi
}%
\providecommand \natexlab [1]{#1}%
\providecommand \enquote  [1]{``#1''}%
\providecommand \bibnamefont  [1]{#1}%
\providecommand \bibfnamefont [1]{#1}%
\providecommand \citenamefont [1]{#1}%
\providecommand \href@noop [0]{\@secondoftwo}%
\providecommand \href [0]{\begingroup \@sanitize@url \@href}%
\providecommand \@href[1]{\@@startlink{#1}\@@href}%
\providecommand \@@href[1]{\endgroup#1\@@endlink}%
\providecommand \@sanitize@url [0]{\catcode `\\12\catcode `\$12\catcode
  `\&12\catcode `\#12\catcode `\^12\catcode `\_12\catcode `\%12\relax}%
\providecommand \@@startlink[1]{}%
\providecommand \@@endlink[0]{}%
\providecommand \url  [0]{\begingroup\@sanitize@url \@url }%
\providecommand \@url [1]{\endgroup\@href {#1}{\urlprefix }}%
\providecommand \urlprefix  [0]{URL }%
\providecommand \Eprint [0]{\href }%
\providecommand \doibase [0]{http://dx.doi.org/}%
\providecommand \selectlanguage [0]{\@gobble}%
\providecommand \bibinfo  [0]{\@secondoftwo}%
\providecommand \bibfield  [0]{\@secondoftwo}%
\providecommand \translation [1]{[#1]}%
\providecommand \BibitemOpen [0]{}%
\providecommand \bibitemStop [0]{}%
\providecommand \bibitemNoStop [0]{.\EOS\space}%
\providecommand \EOS [0]{\spacefactor3000\relax}%
\providecommand \BibitemShut  [1]{\csname bibitem#1\endcsname}%
\let\auto@bib@innerbib\@empty
\bibitem [{\citenamefont {Drela}\ and\ \citenamefont {Youngren}()}]{avl}%
  \BibitemOpen
  \bibfield  {author} {\bibinfo {author} {\bibfnamefont {M.}~\bibnamefont
  {Drela}}\ and\ \bibinfo {author} {\bibfnamefont {H.}~\bibnamefont
  {Youngren}},\ }\href@noop {} {\enquote {\bibinfo {title} {Avl (3.36) user
  guide, avl overview},}\ }\bibinfo {howpublished}
  {\url{http://web.mit.edu/drela/Public/web/avl/}},\ \bibinfo {note} {accessed:
  2018-07-11}\BibitemShut {NoStop}%
\bibitem [{\citenamefont {Etkin}(1959)}]{Etkin_59}%
  \BibitemOpen
  \bibfield  {author} {\bibinfo {author} {\bibfnamefont {B.}~\bibnamefont
  {Etkin}},\ }\href@noop {} {\emph {\bibinfo {title} {Dynamics of Flight:
  Stability and Control}}}\ (\bibinfo  {publisher} {John Wiley $\&$ Sons,
  Inc},\ \bibinfo {address} {New York},\ \bibinfo {year} {1959})\BibitemShut
  {NoStop}%
\bibitem [{\citenamefont {Etkin}(1972)}]{Etkin}%
  \BibitemOpen
  \bibfield  {author} {\bibinfo {author} {\bibfnamefont {B.}~\bibnamefont
  {Etkin}},\ }\href@noop {} {\emph {\bibinfo {title} {Dynamics of Atmospheric
  Flight}}}\ (\bibinfo  {publisher} {John Wiley $\&$ Sons, Inc},\ \bibinfo
  {address} {New York},\ \bibinfo {year} {1972})\BibitemShut {NoStop}%
\bibitem [{\citenamefont {Carpenter}(1997)}]{Carpenter}%
  \BibitemOpen
  \bibfield  {author} {\bibinfo {author} {\bibfnamefont {C.}~\bibnamefont
  {Carpenter}},\ }\href@noop {} {\emph {\bibinfo {title} {Flightwise Volume
  2-Aircraft Stability And Control}}}\ (\bibinfo  {publisher} {Airlife
  Publishing Ltd},\ \bibinfo {year} {1997})\BibitemShut {NoStop}%
\bibitem [{\citenamefont {Cunis~et al}(2019)}]{Cunis_2019}%
  \BibitemOpen
  \bibfield  {author} {\bibinfo {author} {\bibfnamefont {T.}~\bibnamefont
  {Cunis~et al}},\ }\href@noop {} {\bibfield  {journal} {\bibinfo  {journal}
  {Journal of Aircraft}\ }\textbf {\bibinfo {volume}
  {https://doi.org/10.2514/1.C035455}} (\bibinfo {year} {2019})}\BibitemShut
  {NoStop}%
\bibitem [{\citenamefont {Donald~Christiansen}\ and\ \citenamefont
  {Fink}(2005)}]{transferfunctions}%
  \BibitemOpen
  \bibfield  {author} {\bibinfo {author} {\bibfnamefont {R.~K.~J.}\
  \bibnamefont {Donald~Christiansen}}\ and\ \bibinfo {author} {\bibfnamefont
  {D.~G.}\ \bibnamefont {Fink}},\ }\href@noop {} {\emph {\bibinfo {title} {The
  Electronics Engineers' Handbook}}}\ (\bibinfo  {publisher} {5th Edition
  McGraw-Hill},\ \bibinfo {year} {2005})\ pp.\ \bibinfo {pages}
  {19.1--19.30}\BibitemShut {NoStop}%
\bibitem [{\citenamefont {Paz}(2015)}]{paz2015introduction}%
  \BibitemOpen
  \bibfield  {author} {\bibinfo {author} {\bibfnamefont {S.~P.}\ \bibnamefont
  {Paz}},\ }\href@noop {} {\bibfield  {journal} {\bibinfo  {journal} {Ciencia y
  poder a{\'e}reo}\ }\textbf {\bibinfo {volume} {10}},\ \bibinfo {pages} {39}
  (\bibinfo {year} {2015})}\BibitemShut {NoStop}%
\bibitem [{\citenamefont {Karamcheti}(1966)}]{karamcheti1966principles}%
  \BibitemOpen
  \bibfield  {author} {\bibinfo {author} {\bibfnamefont {K.}~\bibnamefont
  {Karamcheti}},\ }\href@noop {} {\emph {\bibinfo {title} {Principles of
  ideal-fluid aerodynamics}}}\ (\bibinfo  {publisher} {Wiley New York},\
  \bibinfo {year} {1966})\BibitemShut {NoStop}%
\bibitem [{\citenamefont {Budziak}(2015)}]{budziak2015aerodynamic}%
  \BibitemOpen
  \bibfield  {author} {\bibinfo {author} {\bibfnamefont {K.}~\bibnamefont
  {Budziak}},\ }\href@noop {} {\emph {\bibinfo {title} {Aerodynamic Analysis
  with Athena Vortex Lattice (AVL)}}}\ (\bibinfo  {publisher} {Hamburg:
  Aircraft Design and Systems Group (AERO), Department of Automotive~…},\
  \bibinfo {year} {2015})\BibitemShut {NoStop}%
\bibitem [{\citenamefont {Griffiths}(2005)}]{griffiths2005introduction}%
  \BibitemOpen
  \bibfield  {author} {\bibinfo {author} {\bibfnamefont {D.~J.}\ \bibnamefont
  {Griffiths}},\ }\href@noop {} {\emph {\bibinfo {title} {Introduction to
  electrodynamics}}}\ (\bibinfo  {publisher} {AAPT},\ \bibinfo {year}
  {2005})\BibitemShut {NoStop}%
\bibitem [{\citenamefont {Cook}(2007)}]{main}%
  \BibitemOpen
  \bibfield  {author} {\bibinfo {author} {\bibfnamefont {M.~V.}\ \bibnamefont
  {Cook}},\ }\href@noop {} {\emph {\bibinfo {title} {Flight Dynamics
  Principles, A Linear Systems Approach to Aircraft Stability and Control}}}\
  (\bibinfo  {publisher} {Elsevier Ltd},\ \bibinfo {address} {30 Corporate
  Drive, Suite 400, Burlington, MA 01803, USA},\ \bibinfo {year}
  {2007})\BibitemShut {NoStop}%
\bibitem [{\citenamefont {Strogatz}(2015)}]{Strogatz}%
  \BibitemOpen
  \bibfield  {author} {\bibinfo {author} {\bibfnamefont {S.}~\bibnamefont
  {Strogatz}},\ }\href@noop {} {\emph {\bibinfo {title} {Nonlinear Dynamics and
  Chaos}}}\ (\bibinfo  {publisher} {Westview Press},\ \bibinfo {address}
  {Colorado},\ \bibinfo {year} {2015})\BibitemShut {NoStop}%
\bibitem [{\citenamefont {Andronov~et al}(1973)}]{Andronov}%
  \BibitemOpen
  \bibfield  {author} {\bibinfo {author} {\bibfnamefont {A.~A.}\ \bibnamefont
  {Andronov~et al}},\ }\href@noop {} {\emph {\bibinfo {title} {Qualitative
  Theory of Second-Order Dynamic Systems}}}\ (\bibinfo  {publisher} {John Wiley
  $\&$ Sons, Inc},\ \bibinfo {address} {New York},\ \bibinfo {year}
  {1973})\BibitemShut {NoStop}%
\bibitem [{\citenamefont {Ogunwa}\ and\ \citenamefont
  {Abdullah}(2016)}]{Ogunwa_2016}%
  \BibitemOpen
  \bibfield  {author} {\bibinfo {author} {\bibfnamefont {T.~T.}\ \bibnamefont
  {Ogunwa}}\ and\ \bibinfo {author} {\bibfnamefont {E.~J.}\ \bibnamefont
  {Abdullah}},\ }\href {\doibase 10.1088/1757-899x/152/1/012022} {\bibfield
  {journal} {\bibinfo  {journal} {{IOP} Conference Series: Materials Science
  and Engineering}\ }\textbf {\bibinfo {volume} {152}},\ \bibinfo {pages}
  {012022} (\bibinfo {year} {2016})}\BibitemShut {NoStop}%
\bibitem [{\citenamefont {Park}\ \emph {et~al.}(2017)\citenamefont {Park},
  \citenamefont {Choi}, \citenamefont {Jo},\ and\ \citenamefont
  {Choi}}]{Park_2017}%
  \BibitemOpen
  \bibfield  {author} {\bibinfo {author} {\bibfnamefont {J.}~\bibnamefont
  {Park}}, \bibinfo {author} {\bibfnamefont {J.-Y.}\ \bibnamefont {Choi}},
  \bibinfo {author} {\bibfnamefont {Y.}~\bibnamefont {Jo}}, \ and\ \bibinfo
  {author} {\bibfnamefont {S.}~\bibnamefont {Choi}},\ }\href {\doibase
  10.2514/1.C034052} {\bibfield  {journal} {\bibinfo  {journal} {Journal of
  Aircraft}\ }\textbf {\bibinfo {volume} {54}},\ \bibinfo {pages} {1} (\bibinfo
  {year} {2017})}\BibitemShut {NoStop}%
\bibitem [{\citenamefont {Sun}\ and\ \citenamefont {Xiong}(2005)}]{Sun}%
  \BibitemOpen
  \bibfield  {author} {\bibinfo {author} {\bibfnamefont {M.}~\bibnamefont
  {Sun}}\ and\ \bibinfo {author} {\bibfnamefont {Y.}~\bibnamefont {Xiong}},\
  }\href@noop {} {\bibfield  {journal} {\bibinfo  {journal} {The Journal of
  Experimental Biology}\ }\textbf {\bibinfo {volume} {208}},\ \bibinfo {pages}
  {447} (\bibinfo {year} {2005})}\BibitemShut {NoStop}%
\bibitem [{\citenamefont {Osano}\ and\ \citenamefont {Ngwenya}(2020)}]{BB}%
  \BibitemOpen
  \bibfield  {author} {\bibinfo {author} {\bibfnamefont {B.}~\bibnamefont
  {Osano}}\ and\ \bibinfo {author} {\bibfnamefont {B.~A.}\ \bibnamefont
  {Ngwenya}},\ }\href@noop {} {\enquote {\bibinfo {title} {Dynamics flight
  stability; weakly nonlinear sampling},}\ } (\bibinfo {year}
  {2020})\BibitemShut {NoStop}%
\end{thebibliography}%
\end{document}